\documentclass[twocolumn,prl,preprintnumbers,superscriptaddress,showpacs]{revtex4}

\usepackage{graphicx}
\usepackage{amsmath,amssymb}
\usepackage{textcomp}

\makeatletter
\DeclareRobustCommand{\chemical}[1]{%
  {\(\m@th
   \edef\resetfontdimens{\noexpand\)%
       \fontdimen16\textfont2=\the\fontdimen16\textfont2
       \fontdimen17\textfont2=\the\fontdimen17\textfont2\relax}%
   \fontdimen16\textfont2=2.7pt \fontdimen17\textfont2=2.7pt
   \mathrm{#1}%
   \resetfontdimens}}
\DeclareRobustCommand{\bchemical}[1]{%
  {\(\m@th
   \edef\resetfontdimens{\noexpand\)%
       \fontdimen16\textfont2=\the\fontdimen16\textfont2
       \fontdimen17\textfont2=\the\fontdimen17\textfont2\relax}%
   \fontdimen16\textfont2=2.7pt \fontdimen17\textfont2=2.7pt
   \mathbf{#1}%
   \resetfontdimens}}
\makeatother

\newcommand{\casrruo}{\chemical{Ca_{2-x}Sr_xRuO_4}}

\newcommand{\srruo}{\chemical{Sr_2RuO_4}}

\newcommand{\srruoruo}{\chemical{Sr_3Ru_2O_7}}

\newcommand{\caruo}{\chemical{Ca_2RuO_4}}

\newcommand{\caruoef}{\chemical{Ca_{1.5}Sr_{0.5}RuO_4}}

\newcommand{\caruoefz}{\chemical{Ca_{1.48}Sr_{0.52}RuO_4}}

\newcommand{\caruoeda}{\chemical{Ca_{1.38}Sr_{0.62}RuO_4}}

\newcommand{\livo}{\chemical{LiV_2O_4}}

\newcommand{\dxy}{\chemical{d_{xy}}}
\newcommand{\dxz}{\chemical{d_{xz}}}
\newcommand{\dyz}{\chemical{d_{yz}}}
\newcommand{\vq}{\chemical{{\bf q}}}
\newcommand{\qfm}{\chemical{{\bf q_{ic-fm}}}}
\newcommand{\qic}{\chemical{{\bf q_{ic-af}}}}
\newcommand{\vQ}{\chemical{{\bf Q}}}
\newcommand{\vG}{\chemical{{\bf G}}}
\newcommand{\chimac}{\chemical{\chi_{mac}}}

\newcommand{\gband}{\chemical{\gamma}-band}

\newcommand{\kommentar}[1]{}

\begin{document}

\title{Strongly Enhanced Magnetic Fluctuations in a Heavy-mass Layered Ruthenate}

\author{O.~Friedt}%
\affiliation{II. Physikalisches Institut, Universit\"{a}t zu K\"{o}ln, Z\"{u}lpicher
Str. 77, D-50937 K\"{o}ln, Germany}
\affiliation{Laboratoire L\'{e}on Brillouin, C.E.A./C.N.R.S., F-91191-Gif-sur-Yvette CEDEX, France}%

\author{P.~Steffens}
\affiliation{II. Physikalisches Institut, Universit\"{a}t zu K\"{o}ln, Z\"{u}lpicher Str. 77, D-50937 K\"{o}ln, Germany}%

\author{M.~Braden}
\email{braden@ph2.uni-koeln.de}%
\affiliation{II. Physikalisches Institut, Universit\"{a}t zu K\"{o}ln, Z\"{u}lpicher Str. 77, D-50937 K\"{o}ln, Germany}%

\author{Y.~Sidis}%
\affiliation{Laboratoire L\'{e}on Brillouin, C.E.A./C.N.R.S., F-91191-Gif-sur-Yvette CEDEX, France}%

\author{S.~Nakatsuji}
\affiliation{Department of Physics, Kyoto University, Kyoto 606-8502, Japan}

\author{Y.~Maeno}
\affiliation{International Innovation Center and Departement of Physics, Kyoto 606-8502, Japan\\}

\date{\today}

\begin{abstract}
We have studied the magnetic excitations in \casrruo, x=0.52 and
0.62, which exhibit an anomalous high susceptibility and heavy
mass Fermi liquid behavior. Our inelastic neutron scattering
experiments reveal strongly enhanced magnetic fluctuations around
an incommensurate wave vector (0.22,0,0) pointing to a magnetic
instability. The magnetic fluctuations show no correlation in
c-direction and also along the RuO$_2$-planes the signal is
extremely broad, $\Delta q =0.45$\,\AA$^{-1}$. These fluctuations
can quantitatively account for the high specific heat coefficient
and relate to the high macroscopic susceptibility. The magnetic
scattering is attributed to the \gband, the active band for spin
triplet superconductivity in \srruo .
\end{abstract}

\maketitle

The interest in ruthenates initiated by the discovery of the
unconventional superconductivity in \srruo ~\cite{Maeno94} has
revealed a rich variety of physical phenomena.
Inelastic neutron scattering (INS) finds dominating incommensurate
(almost antiferromagnetic) fluctuations arising from strong
nesting in the one-dimensional bands of \dxz - and \dyz
-character \cite{Sidis99,Servant02,Braden03}.
However, there is evidence, that superconductivity in \srruo~is
originating mainly in the two-dimensional band of \dxy -character
\cite{review}. This band exhibits a van-Hove singularity only
50\,meV above the Fermi-level \cite{Singh}, which may be related
to the tendency of the perovskite ruthenates towards
ferromagnetism.

The substitution of Sr by Ca yields a very complex \casrruo-phase
diagram \cite{Nakatsuji00,Nakatsuji0300}. Due to the smaller
ionic radius of Ca compared to that of Sr, the structure first
exhibits a rotation around the c-axis, 1.5$>$x$>$0.5
\cite{Friedt01} followed by a tilt distortion for 0.5$>$x$>$0.2
\cite{Friedt01,Nakatsuji00,Nakatsuji0300}. For the highest
Ca-concentration, x$<$0.2, Mott localization sets in ending in
the antiferromagnetic insulator \caruo~\cite{Braden98}. The
different structural distortions are reflected in the magnetic
properties which could semi-quantitatively be explained through
LDA band structure calculations \cite{Fang01}.

Particularly interesting behavior is found in the concentration
range next to localization but still in the metallic phases
\cite{Nakatsuji0300}, 0.5$>$x$>$0.2, where partial localization
has been proposed \cite{Anisimov}. \caruoef~ exhibits the largest
macroscopic magnetic susceptibility, \chimac, at low temperature,
a factor of 200 higher than the one of pure
\srruo~\cite{Nakatsuji0300}. Furthermore, Nakatsuji et al.
observed magnetic hysteresis \cite{Nakatsuji03} and an
exceptionally large linear coefficient of the specific heat of
255\,mJ/mol-RuK$^2$. This puts \caruoef~ well in the range of
typical heavy fermion compounds and there is only one other
transition metal oxide known with a comparable specific heat
coefficient \cite{Kondo97}.

When searching for the so far undetected ferromagnetic
fluctuations, \caruoef~ seems to be favorable due to its large
magnetic susceptibility. We have studied this material by INS and
find indeed strongly enhanced magnetic fluctuations around the
two-dimensional zone center, which furthermore may account for the
high specific heat ratio.

Several single crystals were obtained with a floating zone
technique \cite{Nakatsuji01}. We have studied two compositions,
x=0.52 and 0.62, with volumes of 140 and 350\,mm$^3$,
respectively. INS experiments were performed on the thermal
triple axis spectrometer 1T at the Orph{\'e}e reactor using
double focusing monochromator and analyzer crystals (pyrolithic
graphite (002)). Since the scattered intensity was rather low
and since it turned out that the magnetic scattering is little
modulated in q-space, we relaxed the diaphragms defining the
scattered beam.  INS measures the imaginary part of the dynamical
susceptibility as function of energy and \vQ :
\begin{equation}\label{eqa}
\frac{d^2 \sigma}{d \Omega d
\omega}=\frac{k_f}{k_i}r^2_0\frac{2F^2(\mathbf{Q})}{\pi g^2
\mu^2_B}\frac{\chi''(\mathbf{Q},\hbar\omega)}{1-\exp(-\hbar\omega/k_BT)}
\end{equation}
where $r^2_0=0.292$\,barn, $g\simeq 2$ is the Land\'e factor, $\mathbf{k}_i$ and
$\mathbf{k}_f$ are the incident and final neutron wave vectors, and $F(\mathbf{Q})$ is
the magnetic form factor.

\begin{figure}
\begin{center}
\includegraphics*[width=\columnwidth]{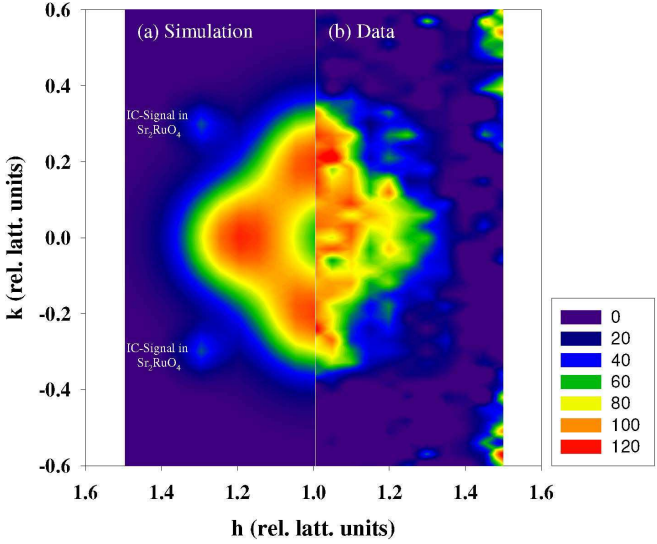}
\caption{Contour plot of the magnetic scattering near
$\mathbf{Q}=(1,0,0)$ in \caruoeda\ at T=12\,K and at an energy
transfer of 4.1\,meV ((a) is a simulation and (b) the measured
data). Due to the enhanced background contribution at lower
scattering angles the scans were restricted to $Q$-values larger
than 1.66\,\AA$^{-1}$. For clarity the background has been
subtracted and the data has been corrected for the magnetic form
factor of \chemical{Ru^{+}}. The schematic representation in (a)
assumes four peaks at $(1\pm0.22,0,0)$ and at $(1,\pm0.22,0)$
with Gaussian profiles and the experimentally determined
parameters, see text.} \label{fig1}
\end{center}
\end{figure}

Most experiments were done with the \caruoeda-crystal due to its
larger volume. Near the \vq -position \qic =(0.3,0.3,$q_l$)
\cite{note-q} where strong incommensurate scattering is observed
in pure \srruo~ \cite{Sidis99} we find only  weak scattering
\cite{Friedtup}, but there is strong scattering around \vQ
=(1,0,0), which is not a zone-center in the three-dimensional
space with the body-centered stacking of the planes. However, due
to the weak inter-layer electronic coupling, magnetism should
have little dependence on the $q_z$-component and we may consider
(1,0,0) as the two-dimensional zone center. Fig.~\ref{fig1} shows
a map of the scattering observed around (1,0,0) at the energy of
4.1\,meV. In contrast to pure \srruo, the magnetic fluctuations
exhibit a broad plateau around the position expected for
ferromagnetism. However, they are still not peaking at the zone
center but remain incommensurate with a very broad maximum at the
finite q-value of \qfm =(0.22,0,0).

We have verified the magnetic nature of the \qfm -scattering by
similar studies around the (3,0,0) and (2,1,0) lattice points
(near (1,1,0) or (2,0,0) phonon scattering is too strong). At the
first positions no similar signal could be detected in agreement
with the expected decrease of a magnetic signal due to the
form-factor. In contrast any phononic signal should get enhanced
at these larger Q-positions. We know, that around (1,0,0) or
equivalent no phonon scattering can appear at energies below
8\,meV \cite{Braden0198}.

\begin{figure}
\begin{center}
\includegraphics*[width=\columnwidth]{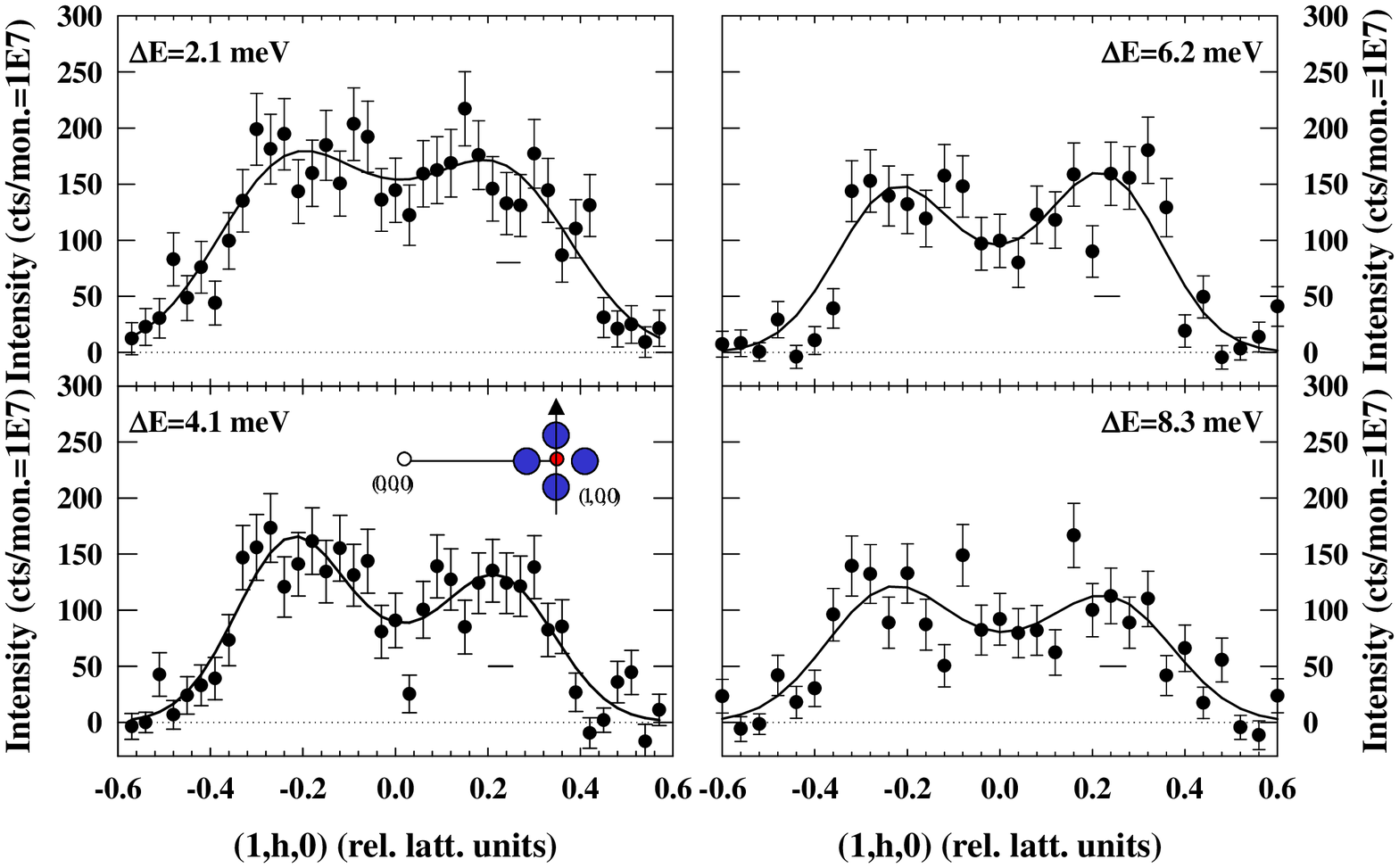}
\caption{Constant-energy scans performed at different energy transfers around
$\mathbf{Q}=(1,0,0)$ along the $[0,1,0]$ direction at T=1.5\,K. The scattering angle
dependent background was subtracted for clarity. The horizontal bars indicate the
spectrometer resolution. The solid lines are fits to Gaussian functions. The data were
acquired with monitor 1E7, which corresponds to a counting time of approximately 7\,min
at 4.1\,meV.} \label{fig2}
\end{center}
\end{figure}

The magnetic scattering in \caruoeda~ is not well defined in \vQ
-space leading to sizeable overlap of the four symmetrically
equivalent contributions from ($\pm$0.22,0,0) and
(0,$\pm$0.22,0). The map as, well as all the scans presented
below, can be described by the superposition of  four Gaussians
isotropic in q-space :
\begin{equation}\label{eqb}
I(\mathbf{Q},\hbar\omega)= BG+\sum_{i=1,...,4}{I_{i}
e^{-\frac{1}{2}(\frac{|{\bf Q} -{\bf Q_i}|}{\sigma })^2}}
\end{equation}

This description allows to analyze the scattering quantitatively,
even though it might be more complex. In the related compound
\srruoruo, for example, a multiple peak structure was observed
\cite{capogna}, which we may not rule out in our case. Figure
\ref{fig2} shows the results of constant energy scans across \vG
=(1,0,0) passing through the two maxima at (1,$\pm$0.22,0) with
fits according to Eq.~(\ref{eqb}). We may follow the
incommensurate scattering corresponding to \qfm~ up to 8\,meV, at
higher energy the contamination with phonon scattering is too
strong. The position of the scattering is independent of the
energy transfer, as expected for a Fermi-surface feature in the
paramagnetic phase.

For the energy of 4.1\,meV at 12\,K we obtain a \vq -width (full
width at half maximum) isotropic in the x,y-plane of 0.45\,\AA
$^{-1}$, which is more than three times larger than the width of
the incommensurate fluctuations in \srruo~\cite{Sidis99}. Thus,
even at this low temperature the fluctuations in \caruoeda~
exhibit a very short correlation length of only about 5\,\AA, but
we may not exclude that the broadening arises from a multiple peak
structure. We have also studied the $q_l$-dependence of this
scattering and did not find any dependency besides the decrease in
intensity with increasing $q_l$ due to the magnetic
\chemical{Ru^+} form factor. The magnetic fluctuations around
\qfm~ are thus not correlated in the $c$-direction similar to the
incommensurate scattering in \srruo~\cite{Servant02}. We find only
a weak increase in the in-plane peak width for decreasing energies
at 1.6\,K, but at higher temperatures, there is significant
additional energy dependent broadening. Astonishingly, the low
energy fluctuations get broadened approximatively when $k_BT$
exceeds their energy \cite{Friedtup}.

\begin{figure}
\begin{center}
\includegraphics*[width=\columnwidth]{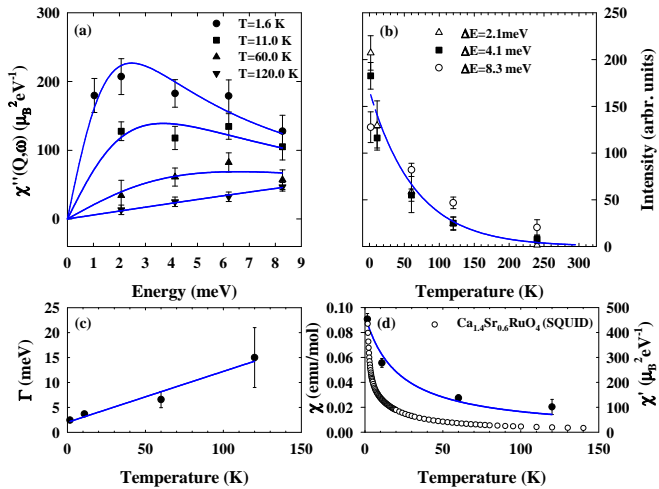}
\caption{(a) : $\chi ''(q_{ic-fm},\omega)$  as function of energy
for different temperatures. The solid lines represent fits with a
single relaxor, Eq.~(3). The data was converted to absolute units
by comparing with data reported for \srruo \cite{Sidis99}. (b)
Temperature dependence of $\chi ''(q_{ic-fm},\omega)$. Temperature
dependence of the characteristic energy (c) and amplitude (d) of
the scattering at \qfm ; in d) we also show the temperature
dependence of \chimac , see \cite{Nakatsuji0300}.} \label{fig3}
\end{center}
\end{figure}

The left part of Fig.~\ref{fig3} gives the energy dependence of
the \qfm-scattering at 1.6\,K which may be described with a single
relaxor fit :
\begin{equation}\label{eqc}
\chi''(\mathbf{q_{ic-fm}},\omega)=\chi'(\mathbf{q_{ic-fm}},0)\frac{\Gamma\omega}{\Gamma^2+\omega^2}
\end{equation}
relating the imaginary part $\chi''(\mathbf{q_{ic-fm}},\omega)$ of
the generalized dynamical susceptibility with the corresponding
real part $\chi'(\mathbf{q_{ic-fm}},0)$ at $\omega=0$. For the
characteristic damping energy we find $\Gamma=2.5\pm0.2$\,meV at
T=1.6\ K, which is much less than the value of 7.5\,meV reported
for the incommensurate spin fluctuations in
\srruo~\cite{Braden03}. We conclude that \caruoeda\ is close to a
magnetic instability, in spite of the rather broad \vQ -width of
the peaks. For the amplitude we obtain
$\chi'(\mathbf{q_{ic-fm}},0)=454\,\mu_B^2eV^{-1}$, which is much
higher than that of the incommensurate scattering in \srruo, which
is indicated in the simulated map in Fig.~\ref{fig1}. Upon heating
the amplitude of the spectrum at \qfm~ decreases and the
characteristic energy shifts to higher values, see Fig.~\ref{fig3}
b)-d). All these observations support the interpretation that
\caruoeda~ approaches a magnetic instability at low temperature.

\begin{figure}
\begin{center}
\includegraphics*[width=0.65\columnwidth]{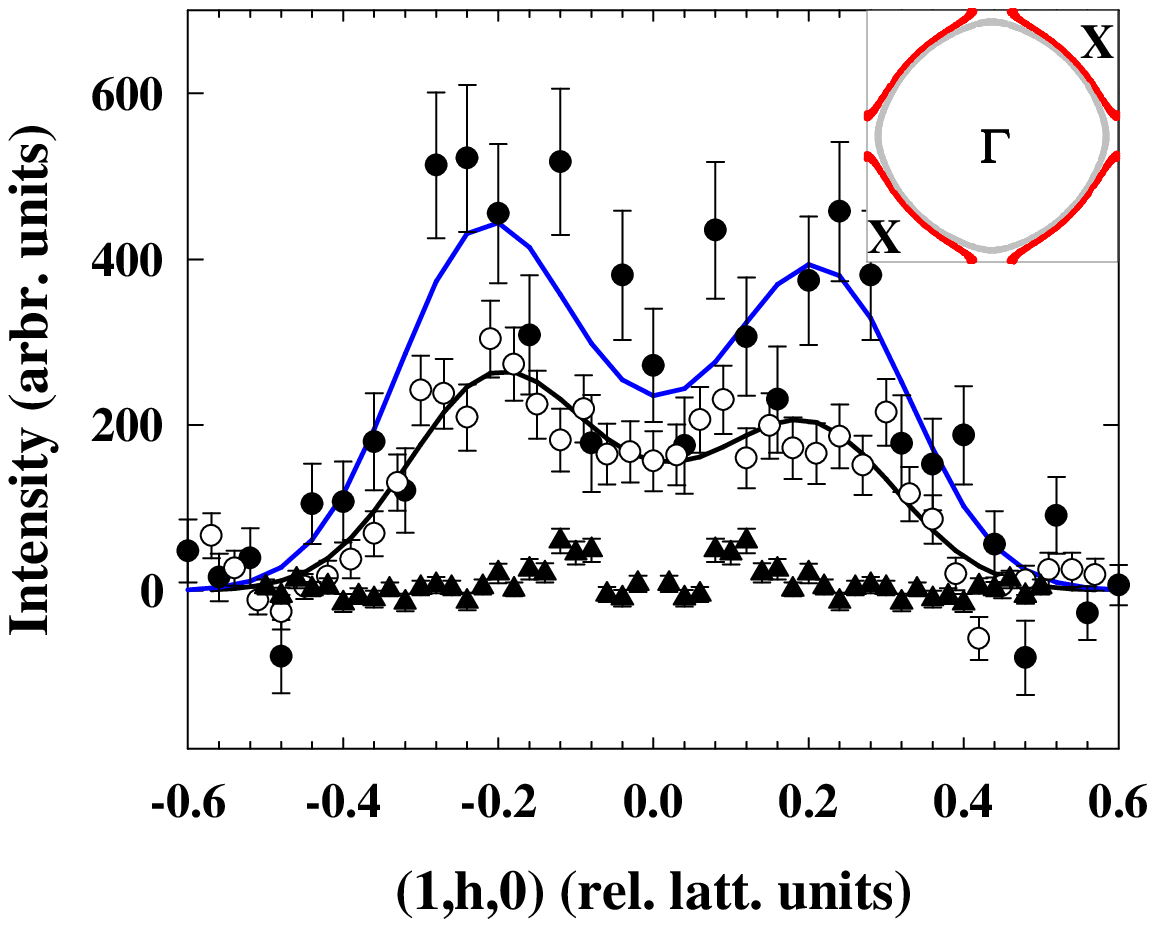}
\caption{(a) Comparison of the inelastic signal around
$\mathbf{Q}=(1,0,0)$ in \casrruo\ for x=2.0
(\protect\includegraphics*[width=0.8em]{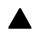}), 0.62
(\protect\includegraphics*[width=0.8em]{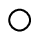}), and 0.52
(\protect\includegraphics*[width=0.8em]{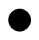}) at an
energy transfer of 4.1\,meV (x=2.0 and 0.62) and 2.1\,meV
(x=0.52). Data was normalized with a reference phonon scan and
background scattering has been subtracted. The data for the pure
compound (x=2.0) was taken from Ref.~\cite{Braden03}. Inset :
sketch of the calculated Fermi surfaces of the $\gamma$ sheet,
indicating the influence of the downward shift of the $d_{xy}$
band due to the rotation of the octahedra.} \label{fig4}
\end{center}
\end{figure}

In Fig.~\ref{fig4} we compare the scattering near \qfm~ for pure
\srruo, \caruoeda~ and \caruoefz. The quasi-ferromagnetic signal is
absent in the pure compound but is even stronger in \caruoefz,
following the behavior of \chimac \cite{Nakatsuji0300}.

The strength of the scattering observed here implies its relevance
for the anomalous physical properties reported for compositions
around \caruoef . The high macroscopic susceptibility relates to
the scattering around \qfm, since, its strong broadening implies a
significant overlap to the ferromagnetic position \vq =(000).
However the temperature dependence of \chimac~ and
$\chi''(q_{ic-fm},\omega)$ or $\chi'(q_{ic-fm},\omega =0)$ are
only qualitatively similar, see Fig.~\ref{fig3}. \chimac~exhibits
a steeper increase at low temperature where it exceeds
$\chi''(q_{ic-fm},\omega)$ by more than a factor five. We conclude
that the anomalously high macroscopic susceptibility evolves from
the little correlated magnetic fluctuations observed here, but the
details of the crossover require further studies at low
temperature and energy. \caruoef~ further exhibits an extremely
high linear coefficient in the specific heat, but it differs from
typical heavy fermion compounds by the strong temperature
dependence of both specific heat coefficient and magnetic
susceptibility, and by the large Wilson ratio of about 40 in the
ruthenate \cite{Nakatsuji03}. There are similarities with the only
other known transition metal compound with comparable C/T-ratio,
\livo, where a magnetic instability and anomalous low temperature
properties were reported \cite{Kondo97,Lee01}. The strongly
enhanced fluctuations in \caruoeda~ can quantitatively account for
the specific heat ratio. Following reference \cite{hayden}, one
may relate the specific heat coefficient with the inverse of the
characteristic energy averaged over the Brillouin-zone: $\gamma =
\frac{\pi k_B^2}{\hbar}\langle{1\over {\Gamma({\bf
Q})}}\rangle_{BZ}$. For \caruoeda, the resulting specific heat
coefficient is large due to the low characteristic energy of
2.5\,meV and due to the broad q-range of the fluctuations. Taking
$\Gamma$ to be constant within a cylinder of radius 0.225\,\AA
$^{-1}$ and neglecting any other contribution we obtain
$\gamma$=250\,mJ/mol-RuK$^2$ in agreement with the direct
measurement. The heavy mass behavior seems thus to arise from the
over-damped magnetic excitations. It is interesting to perform the
same analysis for \srruo: the incommensurate scattering can
account only for about a quarter of the observed
$\gamma$-coefficient suggesting an additional source of
excitations as it has also been deduced from the comparison with
NMR results \cite{Sidis99,Braden03}. It is tempting to assume that
parts of the magnetic fluctuations dominating in \caruoeda, still
play a role for the superconductivity of \srruo where the \gband~
is considered to drive the unconventional superconductivity
\cite{review}.

At temperatures below 1\,K a ferromagnetic ordering has been
reported \cite{Nakatsuji03}, but the ordered moment per Ru appears
to be rather small, of the order of 0.01\,$\mu_B$, in view of the
sizeable magnetic fluctuations seen in our experiment. It is
unclear whether the weak ferromagnetic order reflects the main
magnetic contribution, or whether it arises from some disorder.
We have searched for magnetic ordering in \caruoefz~ down to
0.3\,K. There is no magnetic ordering corresponding to \qfm,
\qic~ or to antiferromagnetism with an ordered moment higher than
0.03\,$\mu_B$.

In the following we want to discuss a possible origin of the
incommensurate scattering around \qfm. A polarized neutron
diffraction study in \caruoef ~ has found predominant
\dxy-character suggesting that the quasi-ferromagnetic instability
is associated with the \gband~\cite{Gukasov02}.  In the right part
of Fig.~\ref{fig4} we show the cylindrical $\gamma$ Fermi-surface
of pure \srruo~ where the van Hove singularity lies near $M$=(0.5
0 0). There has been considerable controversy whether the van Hove
singularity is occupied or not in \srruo~ since the initial ARPES
experiments did not agree with LDA calculations and de Haas van
Alphen measurements. The disagreement was solved when a surface
reconstruction has been found by LEED measurements
\cite{Matzdorf}: in the surface layers, \srruo~ exhibits the same
octahedron rotation around the $c$-axis as the bulk in the samples
studied here, rotation angle of 8.5\textdegree\ compared to the
value of 12\textdegree\ found in the bulk of \caruoef~
\cite{Friedt01}. The structural surface reconstruction induces an
electronic surface state which one may relate to a shift of the
van-Hove singularity below the Fermi-level \cite{Matzdorf}. Since
the rotation distortion is even stronger in \caruoef~ a similar
effect may occur changing the \gband~ from electron-like into
hole-like. Hall effect measurements support this interpretation
\cite{Galvin01}. Using the tight binding parameters of reference
\cite{Singh} but with a down shift of the \gband~ by 100\,meV
taken from the LDA calculation in Ref.~\cite{Fang01} one obtains
the Fermi-surface presented in the inset of Fig.~\ref{fig4} and
indeed one may find a nesting-like vector near (0.2,0,0)
connecting the two ends of hole-like pockets across the $M$-point.
More detailed calculations are required to confirm such
interpretation; in particular the role of band folding induced by
the rotational distortion should be explored. A
Dzyaloshinski-Moriya interaction might also cause a shift of
magnetic fluctuations from the ferromagnetic to an incommensurate
position; however, a rather strong interaction would be needed to
explain the observed \qfm .

In summary we have observed strongly enhanced magnetic excitations
in the heavy mass material \casrruo , x=0.52 and 0.62. In
contrast to pure \srruo , dominant scattering is found around the
zone-center, indicating a quasi-ferromagnetic instability, though
the wave-vector is still finite, \qfm =(0.22,0, 0).  A
description with four symmetrically equivalent broad peaks at
($\pm$0.22,0,0) and (0,$\pm$0.22,0) yields overlap with strong
weight at the zone center corresponding to the ferromagnetic
instability. The enhanced low temperature magnetic susceptibility
in \caruoef~ is thus related with this magnetic instability.
Quantitatively, the strongly enhanced fluctuations account for the
exceptionally high specific heat C/T-ratio. These results shed
further light on the possible role of the \gband~ in
ferromagnetism and in the pairing mechanism  of \srruo.

We acknowledge discussions with P. Pfeuty. This work was supported
by the Deutsche Forschungsgemeinschaft through the
Sonderforschungsbereich 608.

\begin{thebibliography}{28}
\bibitem{note-q}
We use reduced lattice units referring to the undistorted cell  in
\srruo,  $2\pi/a=1.68$\,\AA$^{-1}$ and $2\pi/c=0.50$\,\AA$^{-1}$,
but note that the crystal structure in \caruoeda ~ is heavily
distorted, space group I4$_1$/acd \cite{Friedt01} .
\end{thebibliography}

\end{document}